\documentclass[aps,prb,twocolumn,groupedaddress,superscriptaddress]{revtex4}

\usepackage{amsmath}
\usepackage{graphicx}
\usepackage[FIGTOPCAP]{subfigure}
\usepackage{color}
\usepackage{url}
\usepackage{tikz}
\usepackage{amsfonts}
\usepackage{amssymb}
\usepackage{xcolor}

\begin{document}

\title{Machine learning on the electron-boson mechanism in superconductors}
\author{Wan-Ju Li}
\affiliation{Department of Physics, National Sun Yat-sen University, Kaohsiung 80424, Taiwan}
\affiliation{Institute of Physics, Academia Sinica, Taipei 11529, Taiwan}
\author{Ming-Chien Hsu}
\affiliation{Department of Physics, National Sun Yat-sen University, Kaohsiung 80424, Taiwan}
\author{Shin-Ming Huang}
\email{shinming@mail.nsysu.edu.tw}
\affiliation{Department of Physics, National Sun Yat-sen University, Kaohsiung 80424, Taiwan}
\date{\today}

\begin{abstract}
To unravel pairing mechanism of a superconductor from limited, indirect experimental data is always a difficult task. It is common but sometimes dubious to explain by a theoretical model with some tuning parameters. In this work, we propose that the machine learning might infer pairing mechanism from observables like superconducting gap functions. For superconductivity within the Migdal-Eliashberg theory, we perform supervised learning between superconducting gap functions and electron-boson spectral functions. For simple spectral functions, the neural network can easily capture the correspondence and predict perfectly. For complex spectral functions, an autoencoder is utilized to reduce the complexity of the spectral functions to be compatible to that of the gap functions. After this complexity-reduction process, relevant information of the spectral function is extracted and good performance restores. Our proposed method can extract relevant information from data and can be applied to general function-to-function mappings with asymmetric complexities either in physics or other fields.
\end{abstract}

\maketitle

\section{Introduction}
The mechanism of superconductivity has been one of hottest topics for more than one century since the first experimental evidence of superconductivity in mercury was observed by H. K. Onnes in 1911\cite{Sup1911}.
After half a century, the seminal work by Bardeen, Cooper and Schrieffer \cite{BCS57}, called the BCS theory, elucidating inevitable pairing attraction from the electron-phonon coupling (EPC), 
successfully resolve the long-standing puzzle of the origin of superconductivity.
However, materials with strong EPC could not be incorporated in this scope.
The strong EPC issue in normal metal was solved by Migdal using the field theoretical Green's function approach to show the perturbation series of the EPC can converge quickly due to the negligible vertex correction compared to the self-energy\cite{Migdal58, EPsupC03}. 
The work was further extended to describe the superconducting states by Eliashberg on the basis of collection of states introduced by Bogoliubov \cite{Eliashberg60, Carbotte90}. 
The resulting Eliashberg theory, also known as the Migdal-Eliashberg (ME) theory, is a modification of the BCS theory to take into account the EPC more realistically and thus the retardation effect more accurately \cite{Anderson62}.

The success of EPC for superconductivity demonstrates the huge impact from collaboration between fermions (electrons) and bosons (phonons).
Since then, scientists keep searching for possible bosonic modes behind superconductivity, including unconventional superconductors such as high-temperature superconducting cuprate \cite{cuprates86}
 and iron-based superconductors \cite{FeSC06, FeSC08}. 
Possible bosonic modes include the spin wave, spin fluctuation and so on\cite{bennemann2008}, each of which has specific spectrum.
 In this way, the original BCS theory is widely generalized to describe the superconductivity based on the interactions between electronic and bosonic modes and is applied to various superconducting systems even when the pairing glues of them are expected to be unconventional. Knowing the bosonic spectrum relevant to the superconducting pairing thus gives great insights on the mechanism of superconductivity. 

Theoretically, it is straightforward to calculate gap functions based on the spectrum of bosonic glue of superconductivity. On the contrary, the bosonic spectrum is material dependent and requires case-dependent {\it ab initio} calculations. 
Since electron and boson contribute to the superconductivity through the overall effective electron-boson spectral function (EBSF) without the need to know much material details, the inference of the EBSF from superconducting properties is possible and can stand for a convincing proof of the mechanism.
Our goal is to infer the EBSF from the gap function using machine learning techniques, and this strategy may be extended to other mechanisms including unconventional superconductivity.

Since the beginning of this century, machine learning (ML) related techniques have been intensively developed in the fields of the computer vision and natural language processing\cite{lecun2015,dey2016,minar2018,shrestha2019}. Among the methods for machine learning, the supervised learning and unsupervised learning have been largely applied to various fields other than the computer sciences. In the supervised learning (SL), data with labels are used to train the machine such that the well-trained machine can predict correct labels of data unseen before. For instance, if we have two functions $F_A$ and $F_B$ having some unknown one-to-one correspondence to each other we can train a SL machine to predict $F_B$ for given $F_A$ or vice versa. In the unsupervised learning (USL), a machine is trained by feeding data without labels to learn the distributions of the data with or without the dimension reduction. For instance, the restricted Boltzmann machines can learn the distributions of data without the dimension reduction\cite{montufar2018} while autoencoders (AEs) can learn the distributions in the latent space for a compressed representation in reduced dimensions\cite{dong2018}. 

Starting from few years ago, modern ML gradually becomes a relevant skill for solving different types of problems in physics\cite{carleo2019,dassarma2019,mehta2019,arsenault2014,torlai2016,bezanilla2014,faber2016}. Recently, the detection of phase transitions starts to employ ML as an alternative method \cite{ohtsuki2016,schindler2017,schoenholz2016,wang2016,wetzel2017,chng2018,carrasquilla2017,
vannieuwenburg2017,chng2017}. Since then, many following efforts have been devoted to applying ML to phase transitions of diverse theoretical models \cite{tanaka2017,morningstar2018,heumbeli2018,liu2018,wetzel2017,zhang2017,
zhang2017a,nieuwenberg2018,rem2018,zhang2018,sun2018}. In addition to the phase transition detection, ML techiques can also be applied to various fields, such as guiding the materials discovery\cite{zhou2018,tshitoyan2019,zhang2019,ma2020}, preparing special quantum states \cite{melnikov2018,bukov2018,arrazola2019,bukov2018a,nichols2019} and extracting information from measurements in X-ray experiments\cite{timoshenko2018,carbone2019,carbone2020}. 

For superconductivity, efforts are mainly devoted to the prediction of the transition temperature. These include the similarity search using the encoded fingerprints\cite{isayev2015}, learning the representations of elements in the periodic table to predict superconductivity\cite{konno2018,li2020} and the estimation of the transition temperature via disparate models, such as the support vector machines\cite{owolabi2016,liu2018a}, gradient boosted model\cite{hamidieh2018}, regression models\cite{stanev2018,matsumoto2019}, self-learned descriptors combined with the atom table
convolutional neural networks (ATCNN)\cite{zeng2019}, equation-based model\cite{xie2019}, convolutional gradient boosting decision trees\cite{dan2020}, variational Bayesian neural network\cite{le2020} and models using the text-mining-generated training datasets\cite{court2020}. Effects of training data and improved measures are also investigated for materials discovery \cite{meredig2018}. However, compared to the prediction of the superconducting transition temperatures, the mechanism of the superconductivity attracts less attention. One representative work is the estimation of normal and anomalous self-energies of superconductors from experimental ARPES data\cite{yamaji2019}. 

\begin{figure}[t]
\centering
\includegraphics[width=.9\columnwidth]{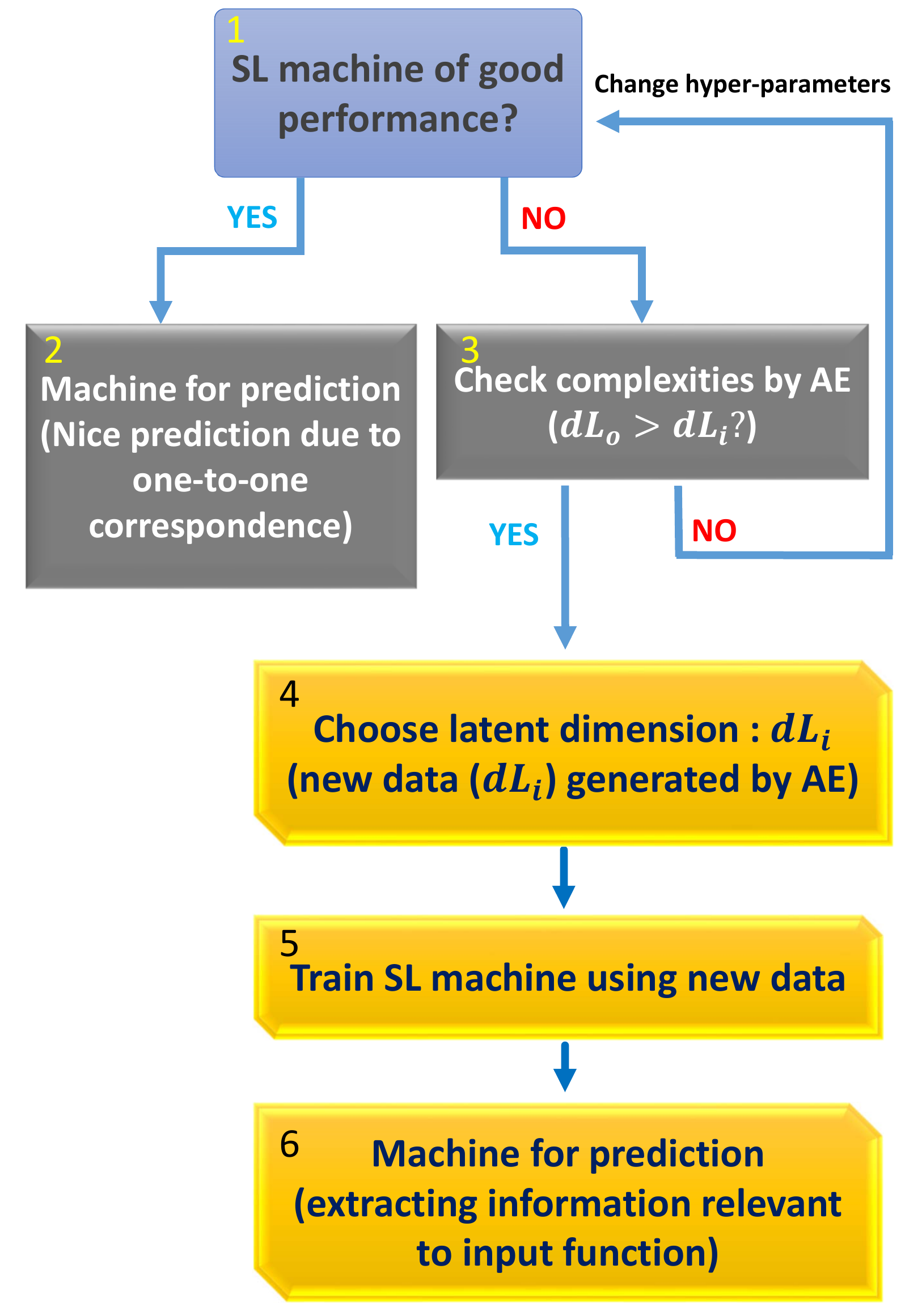}
\caption{The flowchart of the training process. We start at step 1 by training a supervised learning (SL) machine with a dataset. If the performance is good then we can use the well-trained machine for predictions (step 2). If the performance is not good, we have to check if the latent dimension of output functions, $dL_o$, is larger than that of input functions,$dL_i$ (step 3). If it is not the case, we expect the performance can be improved by changing the hyper-parameters of the neural network, such as the number of layers and the number of neurons in each layer (go back to step 1). See also the discussion section. If $dL_o>dL_i$, we firstly train an autoencoder (AE) with latent dimension $dL_i$ using the original dataset and generate new datasets from the well-trained AE (step 4). At step 5, we train a new SL machine with the newly generated dataset and the good performance is expected (step 6).}
\label{fc}
\end{figure}

In this work, we realize a machine which can answer the possible EBSF from the input: a gap function. Through the establishment of the relations learned by the machines, we can derive the spectral function relevant to the gap function without performing the {\it ab initio} calculations.
The performance for simple EBSFs is good by the SL and what is learned by the machine during the training process is investigated.
The SL alone, however, is hard to improve the performance of complex, randomly-generated EBSFs.
In order to resolve this issue, we used the AE to evaluate the complexities of both the EBSFs and gap functions.
It was found that EBSFs have much higher complexity.
Therefore, the complexity of EBSFs is reduced by AE to be compatible to that of the gap function, in which way smoothed EBSFs are generated to contain key information relevant to the input data. 
By using the AE-smoothed EBSFs as the new labels and the AE-transformed gap functions as the new inputs, the performance is greatly improved, which indicates that the new EBSFs preserve the essential information relevant to the new gap function so that they have one-to-one correspondence between them. 
This approach thus proposed a process of complexity reduction which can generally extract the key information of the data relevant to our input functions, and can be utilized to improve SL tasks in physics or other fields. The process is demonstrated in Fig. \ref{fc}.
Note that we also study the cases where the density of state (DOS) is the input. The results are similar to those for the gap functions so only results for the gap functions are presented unless otherwise specified.

The outline of the rest of this article is given in the following. In Sec. \ref{me}, we introduce the ME formalism, describing the relations between the EBSFs and the gap functions, the training datasets. In Sec. \ref{train}, we describe the details and results for the training with simple (Sec. \ref{simple_train}) and complex EBSFs (Sec. \ref{complex_train}). We discuss the relation between the performance and the compatibility of complexity as well as potential applications of our work and then make a conclusion in Sec. \ref{discuss_conclude}.

\section{Migdal-Eliashberg formalism}
\label{me}
The key feature of the superconductivity relies on the Cooper-pair condensation, initiated by the pair of states $(\mathbf{k}\uparrow,  -\mathbf{k}\downarrow)$ occupied coherently. 
Therefore, the Nambu-Gor'kov formalism utilizing the two-component spinor $\psi_\mathbf{k}^\dagger = (c^\dagger_{\mathbf{k}\uparrow}\; c^\dagger_{-\mathbf{k}\downarrow})$ was introduced \cite{Gorkov58, Nambu60, MEwannier13}.
the Green function at momentum k and (imaginary) Matsubara frequency $i \omega_n$ is 
\begin{equation}
[ G(\mathbf{k}, i\omega_n) ]^{-1} = i \omega_n \mathbb{I} - \varepsilon_{\mathbf{k}} \sigma_3 - \Sigma(\mathbf{k}, i\omega_n),
\label{Eq:Dyson}
\end{equation}
where $\epsilon_k$ is the normal-state energy dispersion and $\Sigma$ is the self-energy. For the self-energy $\Sigma$, the ME theory considers the self-energy $\Sigma$ comes from both the EPC and the electron-electron Coulomb interaction.
Especially, the vertex corrections are argued to be small so that a bare vertex is a good approximation, which means the EPC is truncated at order of $\omega_D / E_F$, the ratio between the Debye frequency and the Fermi energy. This leads to the self-energy  
\begin{equation}
\begin{array}{cl}
\Sigma(\mathbf{k}, i\omega_n) &= -\frac{1}{\beta} \sum_{\mathbf{k'}n'\nu} \sigma_3 G(\mathbf{k}', i \omega_{n'})\sigma_3 \\
& \times [ | g_{\mathbf{k}, \mathbf{k}', \nu} |^2 D_{\nu} (\mathbf{k} - \mathbf{k}', i\omega_n - i\omega_{n'}) 
		+ V_C(\mathbf{k} - \mathbf{k}')	],
\end{array}
\label{Eq:sigma_Phonon_Vc}
\end{equation}
where $V_C(\mathbf{k} - \mathbf{k}')$ is the screened Coulomb potential, and $g_{\mathbf{k}, \mathbf{k}', \nu}$ is 
the screened EPC strength which describes scattering 
between electron states $\mathbf{k}$ and $\mathbf{k}'$ through a phonon with wave vector 
$\mathbf{q} = \mathbf{k}' - \mathbf{k}$ and frequency $\omega_{\mathbf{q}\nu}$. 
The propagator for phonons is $D_\nu(\mathbf{q}, i\omega_n) = 2\omega_{\mathbf{q}\nu}/[(i\omega_n)^2 - \omega^2_{\mathbf{q}\nu}]$ and the self-energy is in the form of
\begin{equation}
\Sigma(\mathbf{k}, i\omega_n) = i\omega_n [ 1- Z(\mathbf{k}, i\omega_n)] \mathbb{I} 
		+ \chi(\mathbf{k}, i\omega_n) \sigma_3  
	    + \phi(\mathbf{k}, i\omega_n) \sigma_1 
\label{Eq:SigmaPauli}
\end{equation}
where $Z(\mathbf{k}, i\omega_n)$, $\chi(\mathbf{k}, i\omega_n)$, $\phi(\mathbf{k}, i\omega_n)$ are functions at the Matsubara frequencies to be determined. 
Here, the phase of the pairing potential $\phi(\mathbf{k}, i\omega_n)$ is gauged so as to exclude the $\sigma_2$ component.
Therefore, the Dyson equation becomes
\begin{equation}
[ G(\mathbf{k}, i\omega_n) ]^{-1}  = i \omega_n Z \mathbb{I} 
		-  ( \varepsilon_{\mathbf{k}} + \chi )\sigma_3 
 	    -  \phi \sigma_1.
\label{Eq:Ginv_Pauli}
\end{equation}
The functions of $Z(\mathbf{k}, i\omega_n)$, $\chi(\mathbf{k}, i\omega_n)$ and $\phi(\mathbf{k}, i\omega_n)$
can be determined.

The functions depend on $\mathbf{k}$, which impedes heavily on the computational task. A standard approximation is applied that the DOS is constant of $N_F$ around the Fermi energy $E_F$ since $E_F$ is much larger than the pairing energy. Hence the energy shift $\chi$ is a constant to be omitted.
Furthermore, for conventional superconductors the gap function anisotropy is usually weak or smeared out by impurities \cite{MEwannier13}.
Therefore, an isotropic approximation is adopted to average $\mathbf{k}$ over the Fermi surface. 
By defining the electron-phonon (electron-boson) spectral function (EBSF)
\begin{equation}
\alpha^2 F(\omega) = \frac{1}{N_F} \sum_{\mathbf{k}, \mathbf{k}'} \sum_{\nu} 
      |g_{\mathbf{k}, \mathbf{k}', \nu}|^2 \delta(\epsilon_{\mathbf{k}'})
      \delta(\epsilon_{\mathbf{k}}) \delta(\omega - \omega_{\mathbf{q}\nu})
\end{equation}
where the function is positive definite, 
we finally have the ME equations to solve self-consistently in the following
\begin{equation}
Z(i\omega_n) = 1 + \frac{\pi T}{\omega_n} \sum_{n'}
		\frac{ \omega_{n'}}{R(i \omega_{n'}) }  \lambda(n' - n)
\label{Eq:Ziw}
\end{equation}
\begin{equation}
Z(i\omega_n) \Delta( i \omega_n) = \pi T \sum_{n'}  
	\frac{  \Delta( i \omega_{n'}) }{R(i \omega_{n'}) } [ \lambda( n' -n ) -  \mu^* \theta( \omega_c - | \omega_{n'} | ) ]
\label{Eq:Zdelta}
\end{equation}
where $R(i \omega_n) \equiv \sqrt{\omega_n^2 + \Delta^2( i \omega_n)}$, $\Delta \equiv \phi/Z$ is the gap function, and 
\begin{equation}
\lambda( n' - n ) =
\int \frac{2\omega \alpha^2 F(\omega)}{(\omega_n - \omega_{n'})^2 + \omega^2} d\omega,
\label{lamb}
\end{equation}
is the interaction kernel function. T is the temperature, and the reduced Planck constant $\hbar$ and the Boltzmann constant $k_B$ are set to unity for convenience.

The electron-electron interaction is renormalized and then considered via the the Morel-Anderson pseudopotential $\mu^*$ in the frequency below $\omega_c$.
The $\mu^*$ value is empirical and in the range of $\ 0.1 \sim 0.2\ $ for most conventional superconductors \cite{MEwannier13}. 
Its value usually does not change the behaviour of the results much.
The summation of the Matsubara frequencies is truncated in the numerical calculations.
Once $Z(i\omega_n)$ and $\Delta(i\omega_n)$ are calculated at the Matsubara frequencies, we perform analytical continuation $i\omega_n \rightarrow \omega^+ = \omega + i\Gamma$, where $\omega$ is the real frequency and $\Gamma \rightarrow 0^+$ an infinitesimal positive number. Numerically this can be performed under Pad{\'e} approximation \cite{Pade}. In Appendix \ref{numerical}, we show the details of the numerical calculations of the ME equations and the preparation of the input (gap function $\Delta(\omega)$) and output (EBSF $\alpha^2 F(\omega)$) part of the training datasets.

\section{Inferring the spectral functions by supervised learning}
\label{train}
\begin{figure}[t]
\centering
\includegraphics[width=.9\columnwidth]{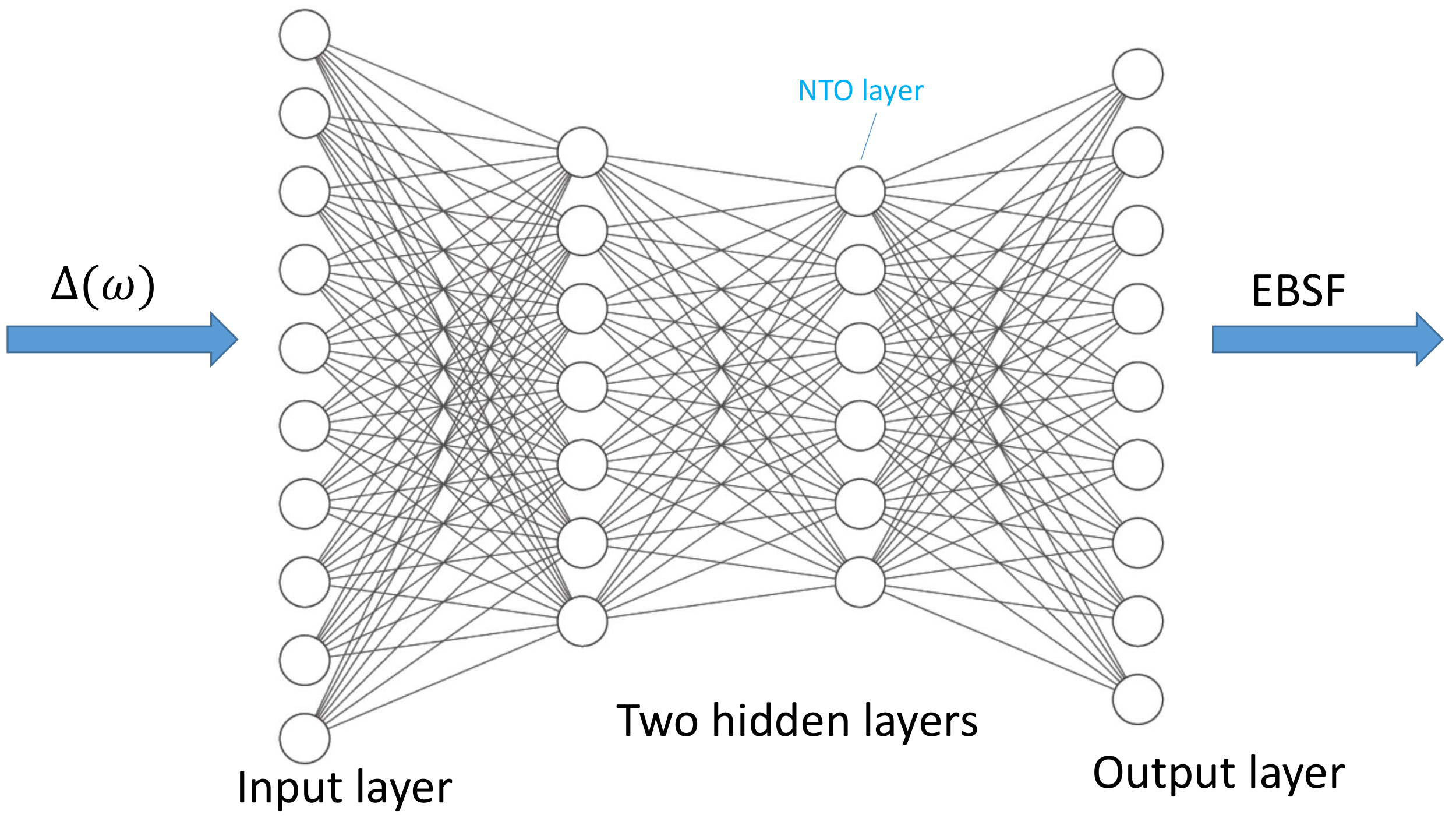}
\caption{The neural network for the supervised learning to predict EBSFs based on the information from the gap functions.}
\label{nn}
\end{figure}

We build up a SL machine by using Keras on the Tensorflow backend to take $\Delta(\omega)$ as an input function to predict the EBSF as the output function. The network structure of the machine contains the input layer (300 neurons), output layer (1000 neurons) and two hidden layers (500 and 800 neurons, respectively), sketched in Fig. \ref{nn}. $\Delta(\omega)$ is fed into the input layer and, after passing through two hidden layers, the predicted EBSFs, $\alpha^2F(\omega)$, are produced in the output layer. The activation functions for the first three layers are ReLU while a linear function is used in the output layer. We choose the mean-square error (MSE) as the loss function to be minimized by the ADAM optimizer\cite{kingma2014}. We divide our dataset into three subsets: $10\%$ of the data for the testing set, $9\%$ for the validation set and the rest for the training set.

\subsection{Good performance for simple bosonic spectrum}
\label{simple_train}
\begin{figure}[t]
\centering
\subfigure[]{
\includegraphics[width=.9\columnwidth]{test.pdf}}
\subfigure[]{
\includegraphics[width=.9\columnwidth]{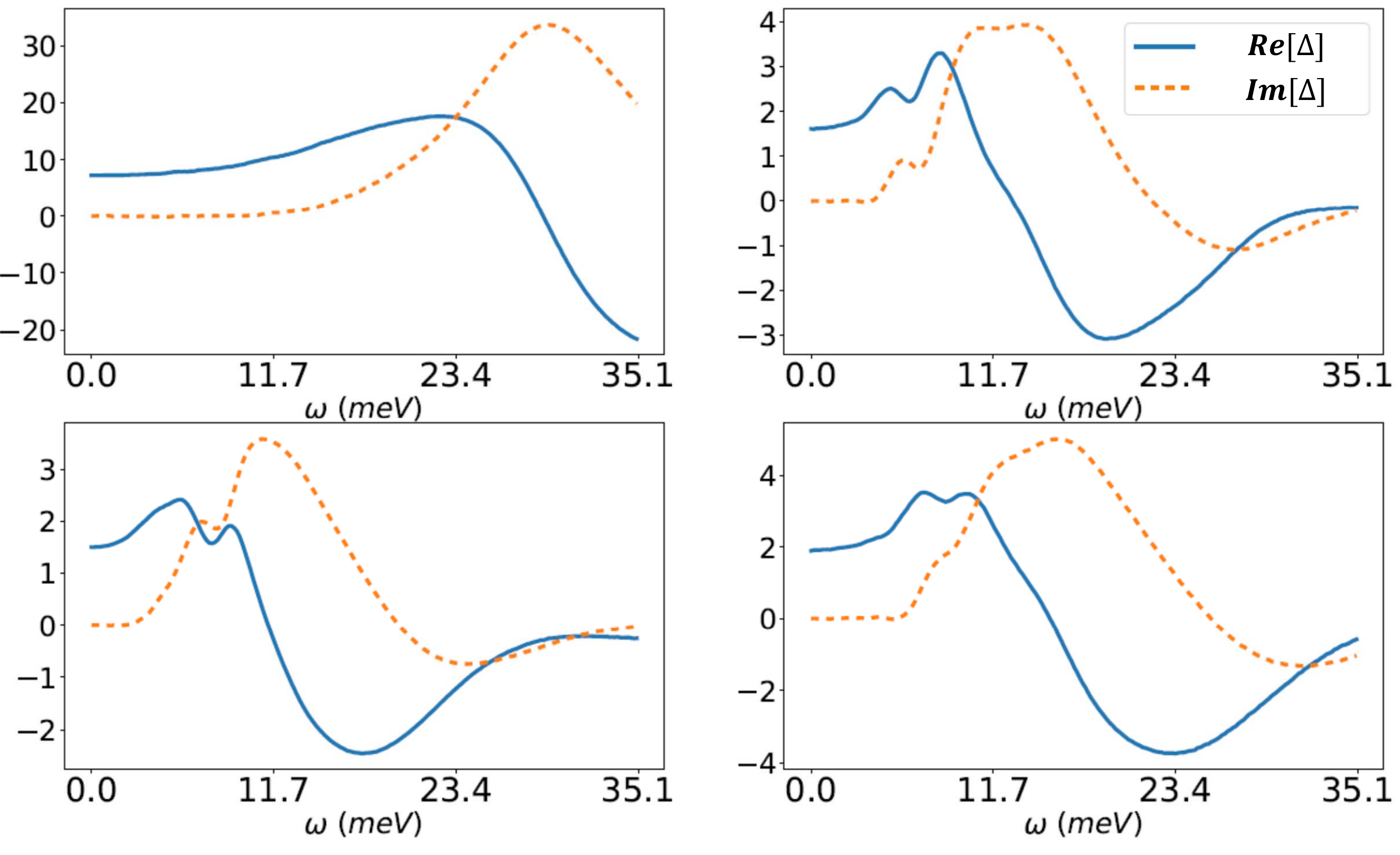}}
\caption{(a). The labels (orange dashed curves) and predictions (blue solid curves) for EBSFs. (b). The real (blue solid curves) and imaginary (orange dashed curves) part of gap functions. }
\label{input}
\end{figure}

In this subsection, we study the performance of the machine when the complexity of the data is low. We prepare the data (totally 1800 data in this dataset) in which every EBSF is constructed by a combination of Gaussian variants. We show some EBSFs in Fig. \ref{input}(a) and gap functions in Fig. \ref{input}(b). Our machine nicely predicts the EBSF, as shown in the close overlap between the real (dashed orange) and predicted (solid blue) functions in Fig. \ref{input}(a). 

Given the good performance, we would like to have insights on what the machine has learned from the training dataset. The details are shown in the appendix \ref{simplelearn}.

\subsection{AE-aided supervised learning for asymmetric complexities}
\label{complex_train}
\begin{figure}[b]
\centering
\includegraphics[width=.9\columnwidth]{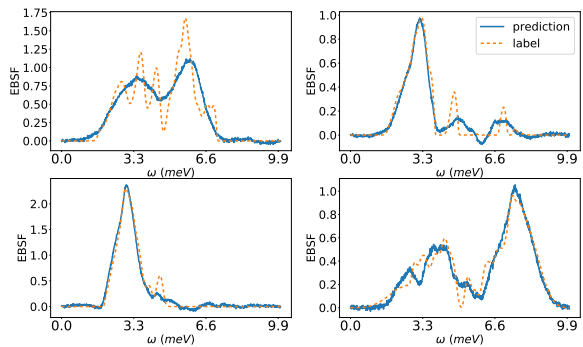}
\caption{The labels (orange dashed curves) and predictions (blue solid curves) for complex EBSFs. The predictions only match the main profile of the labels because the complexity of the EBSFs (output) is much higher than that of the gap functions (input).}
\label{testbf}
\end{figure}

In this part of work, we prepare diverse EBSFs (totally 10000 data in this dataset) by randomly generating the value for each frequency in one EBSF and performing some smoothing (averaging) schemes to slightly eliminate fluctuations. The performance after the training is shown in Fig. \ref{testbf}. The predicted EBSFs (p-EBSFs) match approximately the main profile of the ground-truth EBSFs (gt-EBSFs) but miss fine structures. We propose that this discrepancy is caused by different complexities of EBSFs and gap functions so that they do not have one-to-one correspondence to each other. Explicitly speaking, it is close to a many(EBSFs)-to-one(gap function) mapping, which will be discussed in the appendix \ref{manytoone}. Therefore, in order to restore good performance necessary is the complexity adjustment, during which relevant information of EBSFs can be extracted. In the following, we quantify the complexity by an AE and show explicitly that the complexity of the EBSFs is much higher than that of the gap functions $\Delta(\omega)$.

\begin{figure}[t]
\centering

\includegraphics[width=.9\columnwidth]{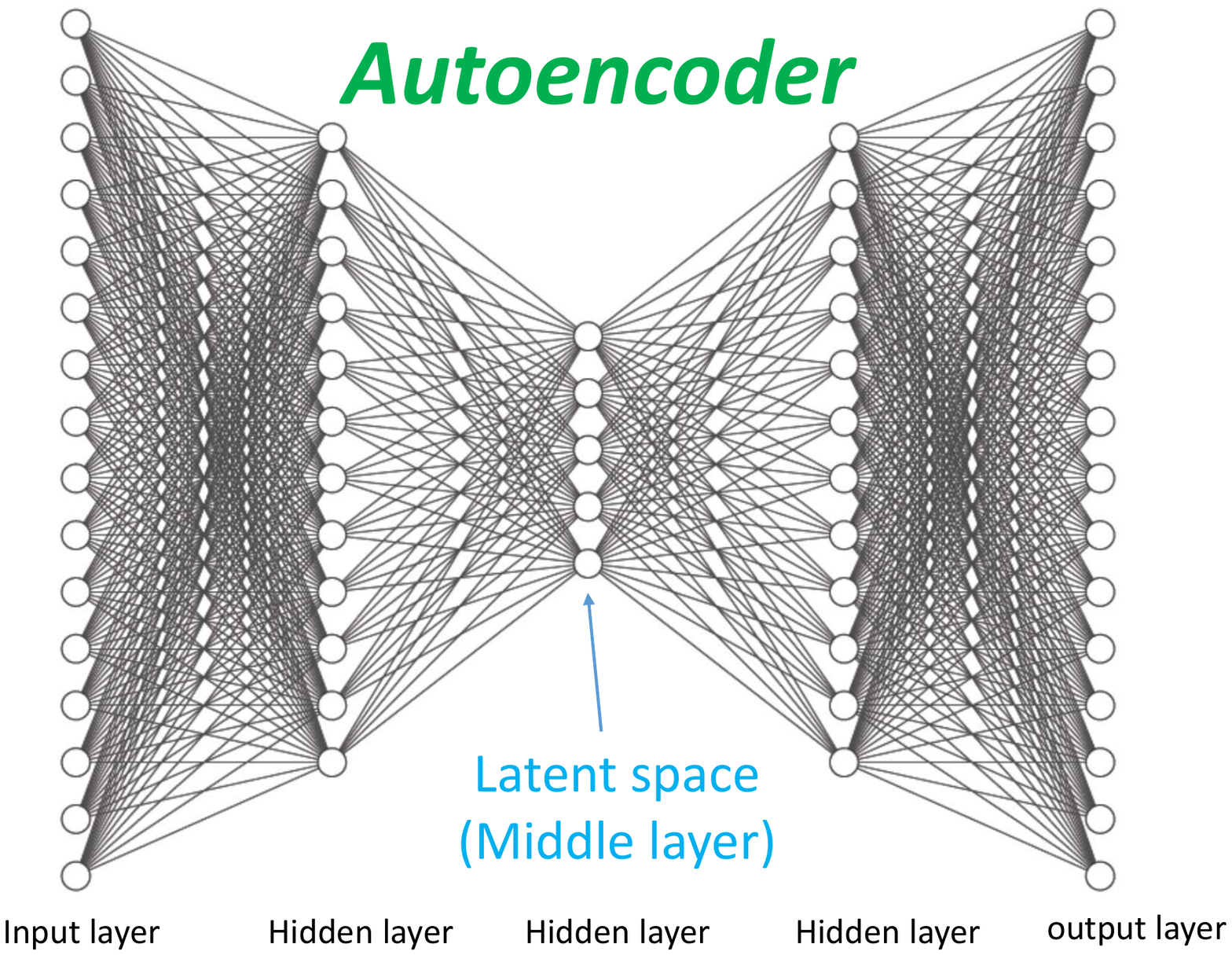}
\caption{The sketch of an autoencoder (AE). An AE contains one encoder and one decoder. An encoder is defined as the part from the input layer to the middle layer (The left three layers.) and a decoder is defined as the part from the middle layer to the output layer (the right three layers). Therefore, an encoder maps a data into an latent vector and a decoder generates a data based on a given latent vector. We train the AE to reconstruct the input in the output layer.}
\label{ae}
\end{figure}

An AE, as illustrated in Fig. \ref{ae}, is usually used to reduce the dimensions of data and to represent data as latent vectors in the latent space. We train an AE to firstly take in a data in the input layer and, after going through the whole AE network, to reconstruct the same data in the output layer. The middle layer of an AE is called the bottleneck whose number of neurons defines the dimension of the latent space, the latent dimension $d_L$. Our strategy is to concatenate each gap function (vector) and the corresponding gt-EBSF (vector) as a new vector and then use those combined vectors to train the AE. 

\begin{figure}[t]
\centering
\subfigure[$\ d_L=1$]{
\includegraphics[width=.3\columnwidth]{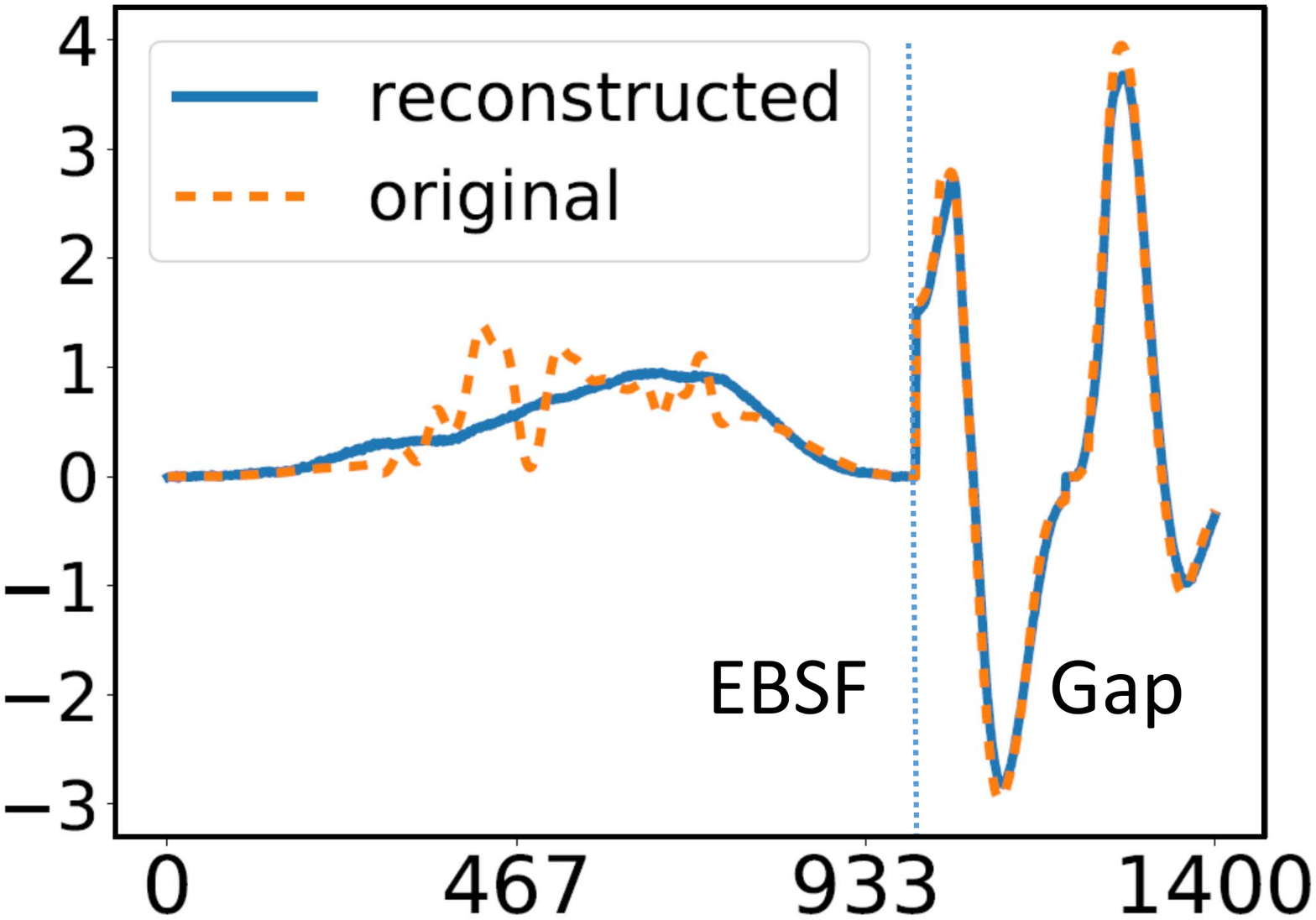}}
\subfigure[$\ d_L=2$]{
\includegraphics[width=.3\columnwidth]{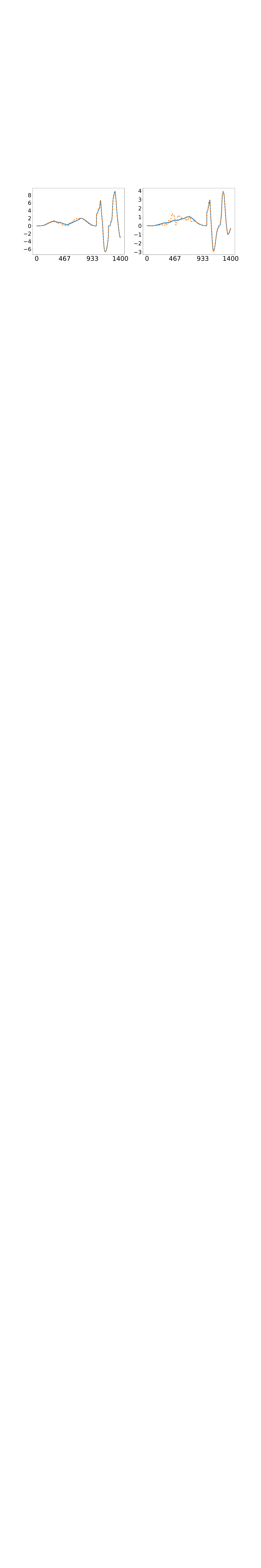}}
\subfigure[$\ d_L=4$]{
\includegraphics[width=.3\columnwidth]{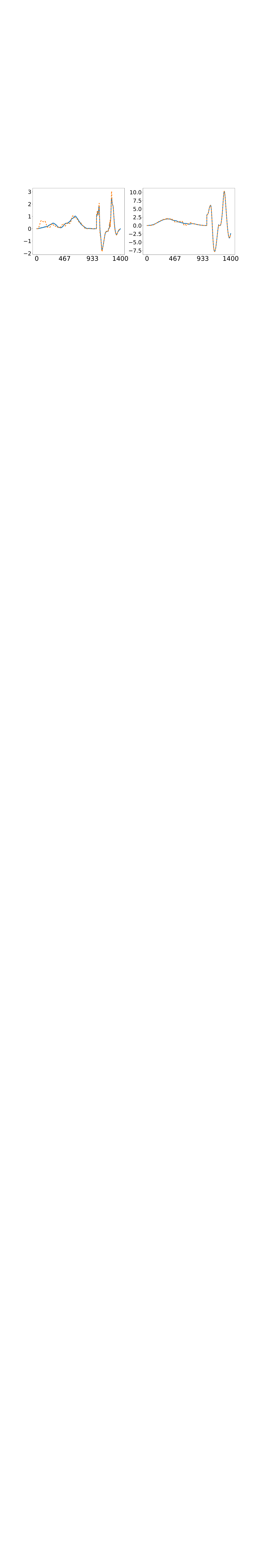}}
\subfigure[$\ d_L=8$]{
\includegraphics[width=.3\columnwidth]{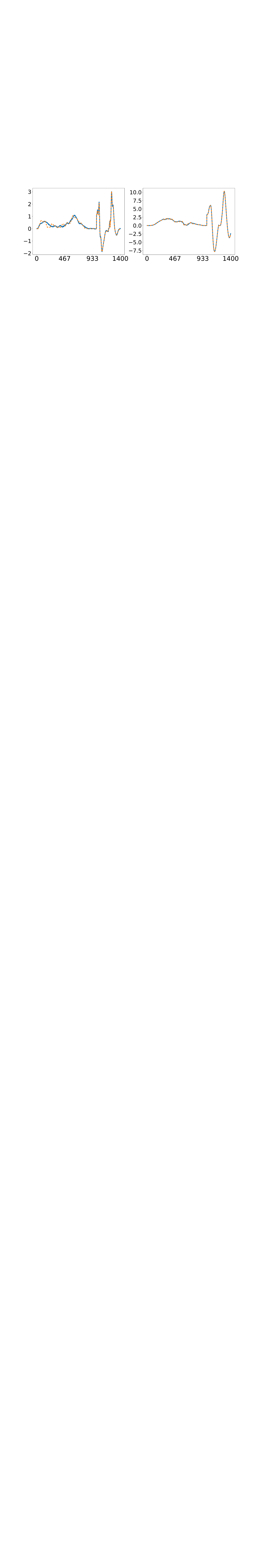}}
\subfigure[$\ d_L=32$]{
\includegraphics[width=.3\columnwidth]{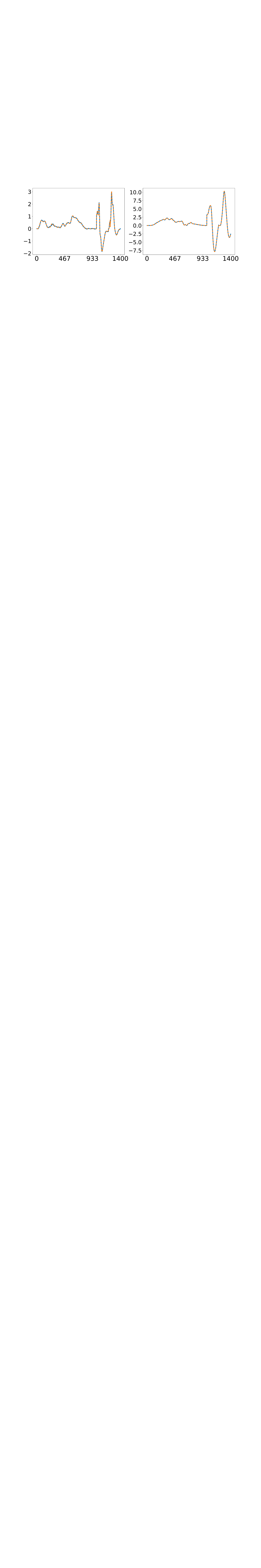}}
\subfigure[$\ d_L=64$]{
\includegraphics[width=.3\columnwidth]{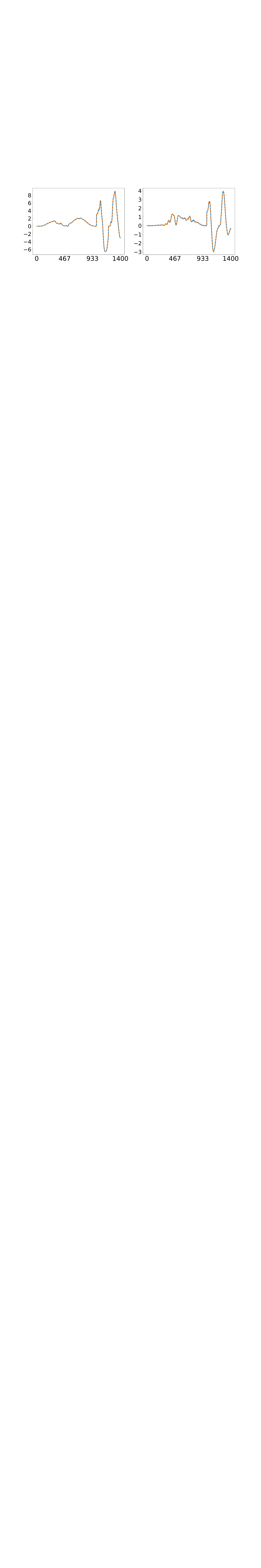}}
\caption{The original (orange dashed curves) and reconstructed (blue solid curves) functions of AE's with different latent dimensions $d_L=1$ (a), $2$ (b), $4$ (c), $8$ (d), $32$ (e) and $64$ (f). The first 1000 components (on the left) of each curve is the EBSF while the last 400 components (on the right) is the gap function. When $d_L\geq 32$, both the EBSF and gap functions can be well reconstructed.}
\label{vaeall}
\end{figure}

As the first trial, we set $d_L=2$ and the result is shown in Fig. \ref{vaeall}(b). This 1400-dimension vector is composed of the EBSF (the first 1000 dimensions) and the gap function (the remaining 400 dimensions), where each dimension is a discrete frequency point of the corresponding functions. The gap function is well reconstructed while the EBSF is not. This result shows that the two dimensional latent space is enough for representing the gap functions but not enough for representing the complex EBSFs. Next, in order to find the required dimension of the latent space for well representing both the EBSFs and gap functions, we study cases with the latent dimensions $d_L=1,2,4,8,32,$ and $64$. The results are shown in Fig. \ref{vaeall}. Both EBSFs and gap functions are well reconstructed for $d_L\geq 32$, which indicates that the latent dimension for the EBSFs is roughly 32 (The minimal required latent dimension should be between 24 and 32 according to our simulations.)

The two complexities (or $d_L$'s) for EBSFs and gap functions are so different that many distinct EBSFs may map onto similar gap functions, consistent with the proposed many-to-one mapping. In this case, the machine has difficulties in learning the mapping and possibly end up with outputting the averaged EBSF. It is consistent with what has been observed in Fig. \ref{testbf} as the prediction curves seem to serve as smoothed functions of the real curves. This also explains our bad training performance in Fig. \ref{testbf}: randomly generated EBSFs contain much information, and most of it is not relevant to the gap function so that the performance of the SL machine is reduced. Questions are raised: can the simplified p-EBSF be a good prediction? It is possible that this simplified p-EBSFs have contained enough information to reproduce all relevant physics in the gap functions and we do not really need the original, complex EBSFs for understanding the mechanism behind the superconductivity. Maybe all we need are those EBSFs having similar complexity to that of the gap functions.

\begin{figure}[t]
\centering
\subfigure[\ p-EBSF]{
\includegraphics[width=.9\columnwidth]{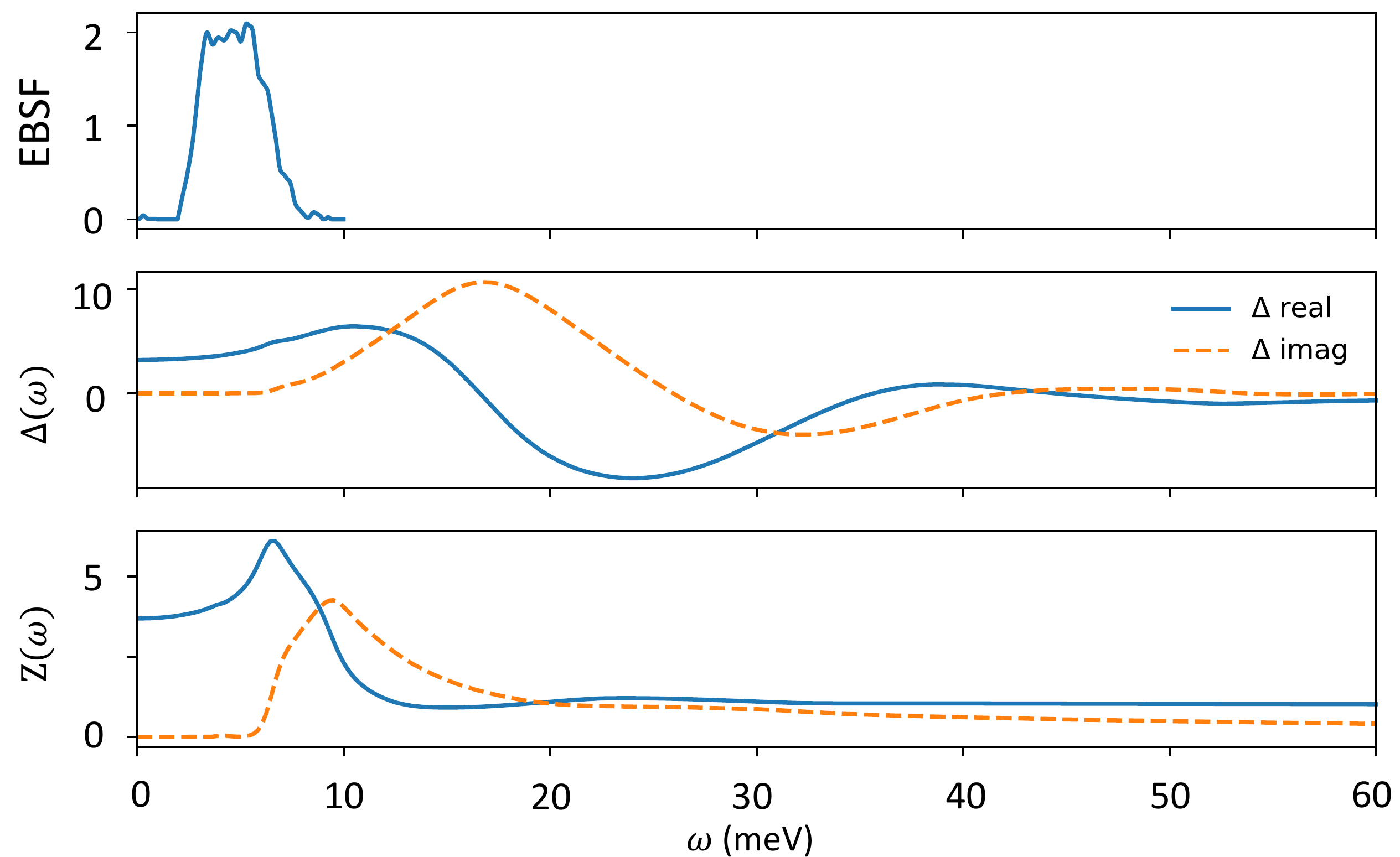}}
\subfigure[\ gt-EBSF]{
\includegraphics[width=.9\columnwidth]{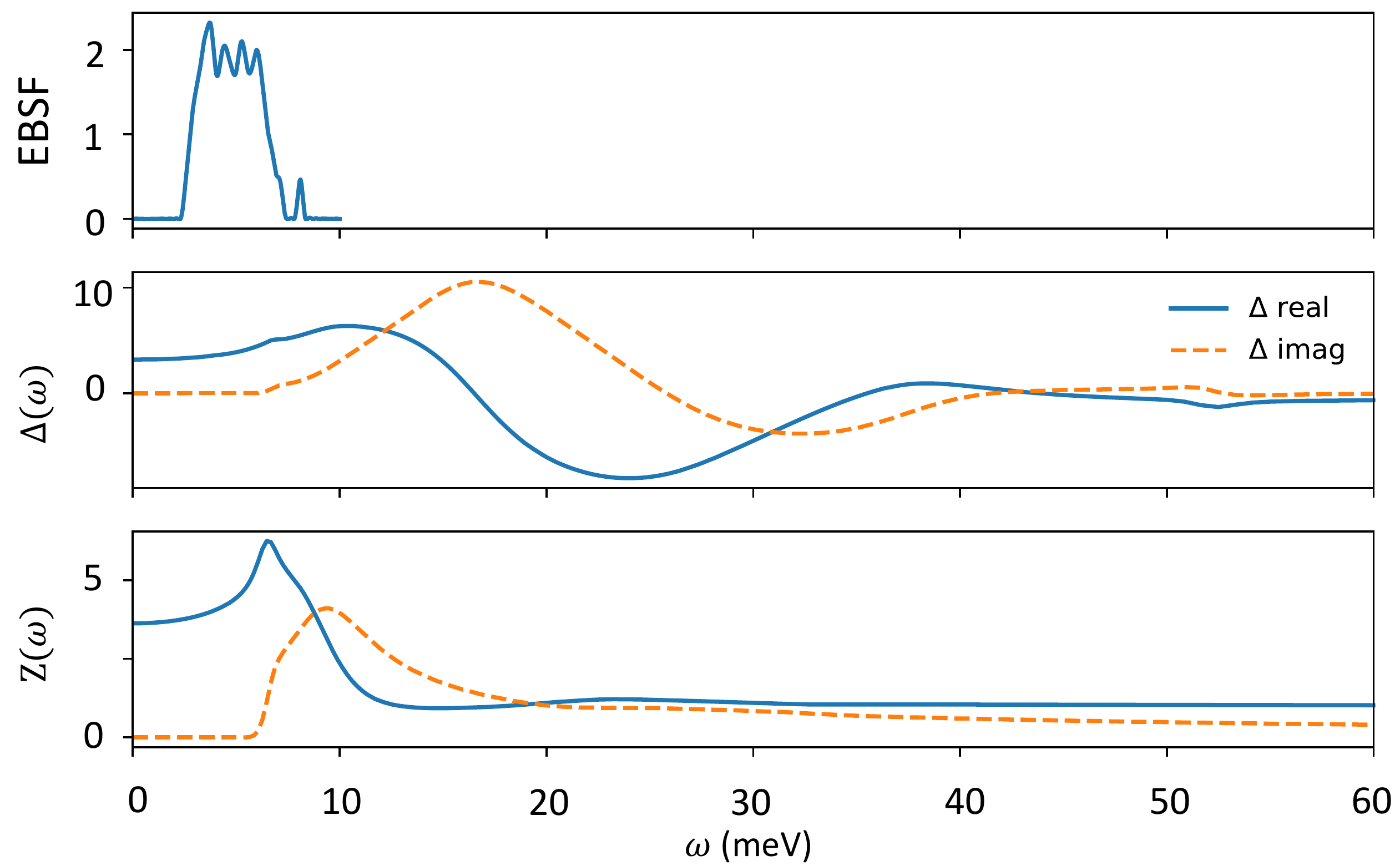}}
\caption{The results based on the p-EBSF (a) and the gt-EBSF (b). The first panel of each subplot is the EBSF and the following two panels are gap-related functions. The difference between two EBSFs is mainly small-scale fluctuations, which turn out to be not physically relevant as the resulting gap-related functions from these two EBSFs are similar. }
\label{pbsgtbs}
\end{figure}

In order to clarify ideas mentioned above, we calculate two gap functions based on the gt-EBSFs and the p-EBSFs. If both the simplified p-EBSF and complex gt-EBSF contain the relevant information, their resulting gap functions should look similar. Our results, as shown in Fig. \ref{pbsgtbs}, supports this idea. For each subfigure, the first panel is the EBSF and the following two panels show the calculated gap-related functions. Although the gt-EBSF has some small-scale fluctuations which do not exist in the p-EBSF, their resulting two functions look similar. It demonstrates that the extra complexity (say, for instance, those small-scale fluctuations) is not physically relevant to the gap functions. Therefore, our SL machine, in fact, does learn the physically relevant part of the EBSFs even though it does not fully reproduce the gt-EBSFs. Given that both the p-EBSFs and gt-EBSFs contain information relevant to the gap functions, we examine what the key information looks like in the following.

\begin{figure}[t]
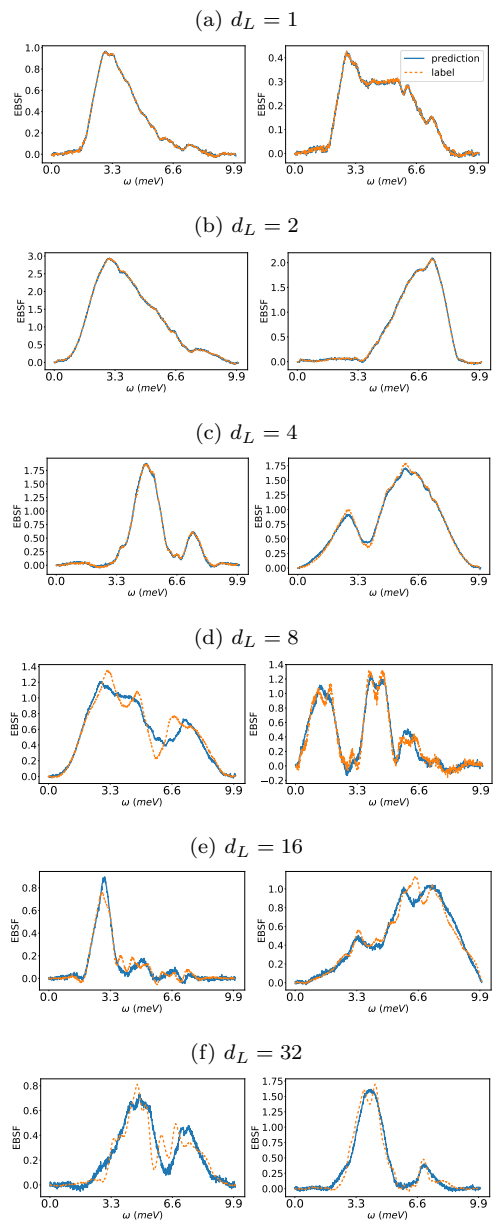

\centering
\subfigure[$\ d_L=1$]{
\includegraphics[width=.75\columnwidth]{aeSL-1.pdf}}
\subfigure[$\ d_L=2$]{
\includegraphics[width=.75\columnwidth]{aeSL-2.pdf}}
\subfigure[$\ d_L=4$]{
\includegraphics[width=.75\columnwidth]{aeSL-4.pdf}}
\subfigure[$\ d_L=8$]{
\includegraphics[width=.75\columnwidth]{aeSL-8.pdf}}
\subfigure[$\ d_L=16$]{
\includegraphics[width=.75\columnwidth]{aeSL-16.pdf}}
\subfigure[$\ d_L=32$]{
\includegraphics[width=.75\columnwidth]{aeSL-32.pdf}}
\caption{The labels (orange dashed curves) and predictions (blue solid curves) for the SL trainings using data generated from the AE with different latent dimensions $d_L=1\sim 32$. The performance of the SL training increases with decreasing $d_L$'s. The difference in the complexities between gap functions and EBSFs decreases with decreasing $d_L$, leading to the improved SL training performance.}
\label{aenewnew}
\end{figure}

Our belief is that if one simplified EBSF do contain mostly the key information relevant to the gap functions, the complexity of that simplified EBSF has to be of the same degree as that of the gap functions. Furthermore, the SL machine, trained by the simplified EBSFs and corresponding gap functions, will restore the good performance. In order to systematically control the complexities, here we use an AE to generate new EBSFs as well as the gap functions with desired complexity (or latent dimension) for our SL trainings. The way to generate new datasets is given in Appendix \ref{newdata}. As an example, we train an AE to generate new dataset $\{v_2\}$ with $d_L=2$ (The blue curve in Fig. \ref{vaeall}(b) is one sample.). The $\{v_2\}$ dataset, containing the gap functions $\{g_2\}$ and EBSFs $\{f_2\}$, is then used to train a new SL machine and the results are shown in Fig. \ref{aenewnew}(b). 

The performance is good for two reasons. Firstly, the latent dimension of the EBSFs is reduced to two, similar to that of the gap functions. The second and most important reason is that the EBSFs and gap functions are co-transformed so that relevant connection between them are kept by the AE. If these two functions are separately transformed by the AE the performance is lowered. In other words, via an AE, the latent dimension of the EBSFs is reduced from 32 to 2 and the co-transformation with the gap functions makes the 2-D latent space of the EBSFs equal to the 2-D latent space of the gap functions. We also check that the AE-transformed EBSFs really lead to, by the self-consistent calculations, the new, co-transformed gap functions for any latent dimension. That is, an AE can preserve the connection between the gap functions and EBSFs and transfer this connection to the new dataset no matter what latent dimension is chosen. In the appendix \ref{relevantproperty}, we provide evidences showing that the EBSFs predicted by our AE-follow-up SL machine do contain information relevant to the gap functions for any latent dimensions. In the appendix \ref{manytoone}, we use the connection preserved by the AE to numerically show the many-to-one mapping between EBSFs and gap functions claimed above.

Given that all newly generated datasets $\{v_n's\}$ for any latent dimensions $n$ satisfy the ME equations, we have to determine which $d_L$ is the appropriate one to construct the SL machine for predictions. The key is to find the latent dimension of the gap functions $d_{L_g}$. From the inspection of Fig. \ref{vaeall}, it is easy to find that $d_{L_g}=2$, which is the minimal dimension where the AE-transformed gap functions $g_2$ look similar to the original gap functions $g_0$. In other words, the phase space of $\{g_2\}$, the collection of gap functions generated by a trained AE with $d_L=2$, may contain most of the phase space of $\{g_0\}$. On the contrary, $\{g_1\}$ may contain a smaller portion of the phase space of $\{g_0\}$, which can be seen, for instance, in Fig. \ref{vaeall}(a) as the reconstructed gap function has some mismatch with the original one. In Appendix \ref{properlatent}, we evaluate the portion of the $\{g_0\}$ phase space occupied by $\{g_n\}$ with several different $n$'s to quantitatively determine $d_{L_g}=2$.

For the cases using the DOS as input functions, similar calculations and analysis have been performed and the appropriate latent dimension is one, instead of two for the gap functions. It is consistent with our observations as the difference between DOS curves is mainly the positions of the coherence peaks while the difference between gap functions can exist in much more different ways. 

Given the good performance shown above, we can summarize our process in Fig. \ref{fc} for improving the performance of a general functional-mapping SL as following. First of all, we check the latent dimensions (or complexities) of both input and output functions $d_{L_i}$ and $d_{L_o}$ respectively by AEs with different $d_L$'s. Usually, the bad performance occurs when $d_{L_o}>d_{L_i}$, which will be discussed next. Secondly we train an AE with the latent dimension $d_{L_i}$. During this second step, the complexities of input and output functions are adjusted to be the same and new dataset can be generated by the trained AE. As the final step, we use newly generated datasets to train the SL machine and use this well-trained machine to make predictions for gap functions from experiments.

\section{Discussions and Conclusions}
\label{discuss_conclude}
The bad performance usually occurs when $d_{L_o}>d_{L_i}$. In this case, the machine has to learn to infer remarkably different output functions (EBSFs in this work) based on similar input functions (gap functions in this work). This is a difficult task and the machine usually ends up with predicting the "coarse grained" functions as mentioned above. However, for the opposite situation, where $d_{L_o}<d_{L_i}$, remarkably different input functions are used to infer similar output functions, which is a relatively easier task.  Consider an extreme example where two different input functions ($q_1$ and $q_2$) correspond to the same output function ($r_1$). Then, the machine only needs to learn to infer $r_1$ no matter which one of $q_1$ and $q_2$ is encountered. For a consistency check, we reverse the roles of input and output functions (that is, here, we use EBSFs to predict gap functions via a SL machine) and the performance of the SL machine is good (not shown), which is consistent with our expectations.  

In order to have good performance, the latent dimension of the AE has to be chosen appropriately if initially the two latent dimensions of input and output functions are largely different. For the simpler case shown in the first part of this work, two complexities (latent dimensions) are compatible to each other so that the performance is good and we do not need an extra AE. For the complex case shown in the second part of this work, the latent dimension of the output functions ($d_{L_o}= 32$) is much larger than that of input functions ($d_{L_i}= 2$ for the gap functions and $d_{L_i}= 1$ for the DOS functions). Therefore the best performance occurs when the latent dimension of the AE is chosen to be two (one) if the gap functions (DOS functions) are chosen as the input functions. 

Usually, it is difficult to have enough dataset directly from experimental inputs for training a SL machine. Therefore, we theoretically generate datasets for training. In order to include all possibilities encountered in the real physical world, data is generated with maximum randomness. However, this process can create extra complexity which is irrelevant to the physical properties of interest leading to bad performance as expected. Our process of complexity reduction serves as a way to extract physically relevant information of the randomly generated functions.

In this work, our method is applied to mainly BCS superconductivity but the extension to other types of order parameters of superconductivity, such as $d$-wave SC or other unconventional ones, should be straightforward as long as similar theoretical formalism connecting the gap functions and bosonic glue is established. Furthermore, this method can be generally applied to any function-to-function supervised-learning tasks with asymmetric complexities to improve the performance and obtain more understanding on the data. 

In summary, we predict the EBSF of a superconductor based on the measured gap functions via the supervised learning aided by the autoencoder. When the latent dimensions of input and output functions are compatible, the performance is good. When these two latent dimensions are remarkably different ($d_{L_o}>d_{L_i}$), we have to reduce the latent dimension of the complex one to be compatible to the simple one so that the good performance restores. This reduction can extract physically relevant information. Therefore, this method paves the ways to improving the performance of general function mappings with asymmetric complexities and to extracting meaningful information in data from physics or other fields.

The authors thank Prof. Ting-Kuo Lee for fruitful discussions. This work is supported by Ministry of Science and Technology (Grant No. MOST 109-2112-M-110-010 and MOST 108-2112-M-110-013-MY3 ) and National Sun Yat-sen University.

\appendix

\section{Many-to-one mapping}
\label{manytoone}
In order to demonstrate the many-to-one mapping, we statistically examine the difference between two EBSFs (say, the original $f_1$ and the complexity-reduced $f_2$) and that between two corresponding gap functions (say, $g_1$ and $g_2$). As the first step we randomly generate $f_1$ and obtain, via ME self-consistent calculations, the gap function $g_1$. We then choose an AE-follow-up SL machine with the latent dimension $d_L=n$ and feed this machine with $g_1$. The output is $f_2$. As the next step, we obtain $g_2$ self-consistently based on $f_2$. Note that, both pairs (($f_1$,$g_1$) and ($f_2$,$g_2$)) are shown in the main text to satisfy the ME equations. This completes the generation of one data ($f_1$, $f_2$, $g_1$, $g_2$) and, in the following, we define and evaluate two differences: $\Delta_f$ for EBSFs and $\Delta_g$ for gap functions using the Pearson correlation coefficient (PCC) \cite{pearson1895}.

For each data, the PCC of $f_1$ and $f_2$ and that of $g_1$ and $g_2$ are calculated as $p_f$ and $p_g$, respectively. The PCC will be equal to 1 if the two compared functions are exactly the same. Then two differences are defined as $\Delta_f=1-p_f$ and $\Delta_g=1-p_g$. We can then calculate the average of $\Delta_f$ and $\Delta_g$ over the dataset for each latent dimension $n$. As shown in table \ref{manyone} that $\Delta_f$ is roughly two orders of magnitude larger than $\Delta_g$, which means remarkably different EBSFs correspond to similar gap functions, a many-to-one mapping.

\begin{table}[ht]
\centering
\caption{The differences in the gap functions ($\Delta_g$) and EBSFs ($\Delta_f$) in terms of the Pearson correlation coefficient (PCC). The difference in EBSFs is much larger than that in gap functions, the many-to-one mapping.}
\begin{tabular}{|c|c|c|c|c|c|}
\hline 
$d_L$ & 2 & 4 & 8 & 16 & 32 \\ 
\hline 
$\Delta_g$ & $0.00336$ & $0.00451$ & $0.0019$ & $0.00234$ & $0.002$ \\ 
\hline 
$\Delta_f$ & $0.107$ & $0.084$ & $0.0624$ & $0.0623$ & $0.0562$ \\ 
\hline
\end{tabular} 
\label{manyone}
\end{table}%

\section{What the machine has learned from simple datasets}
\label{simplelearn}
\begin{figure}[t]
\centering
\subfigure[\ 2 neurons]{
\includegraphics[width=.45\columnwidth]{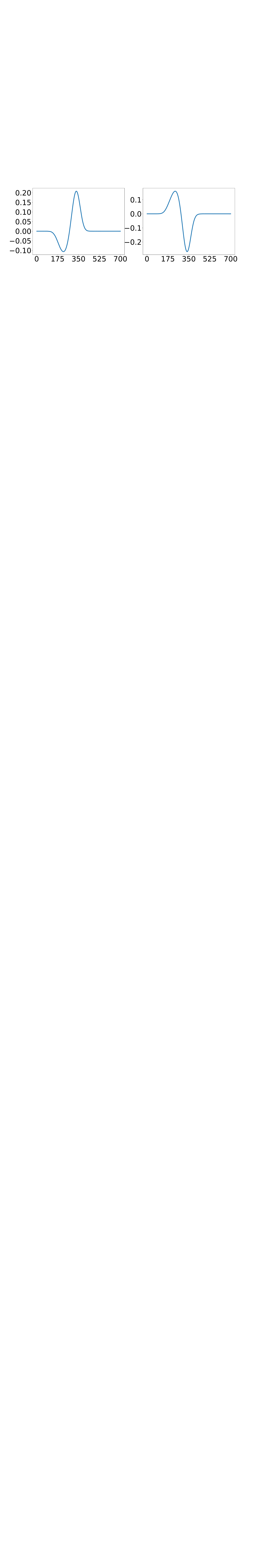}}
\subfigure[\ 4 neurons]{
\includegraphics[width=.45\columnwidth]{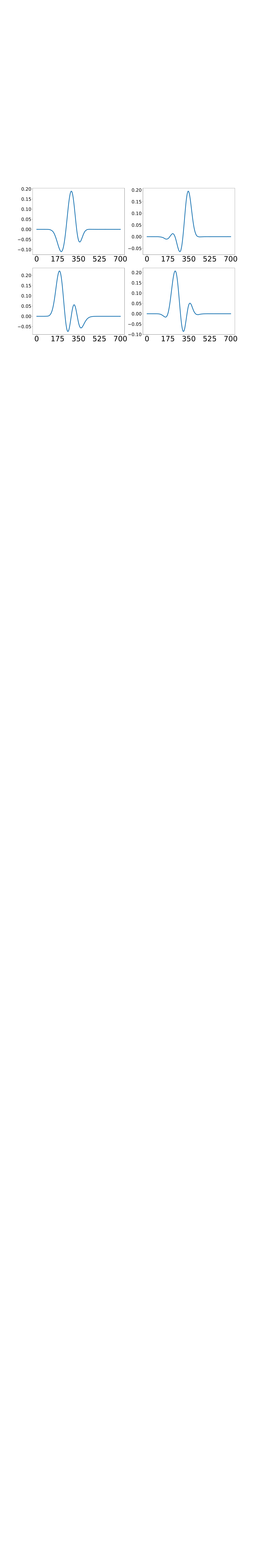}}
\subfigure[\ 6 neurons]{
\includegraphics[width=.45\columnwidth]{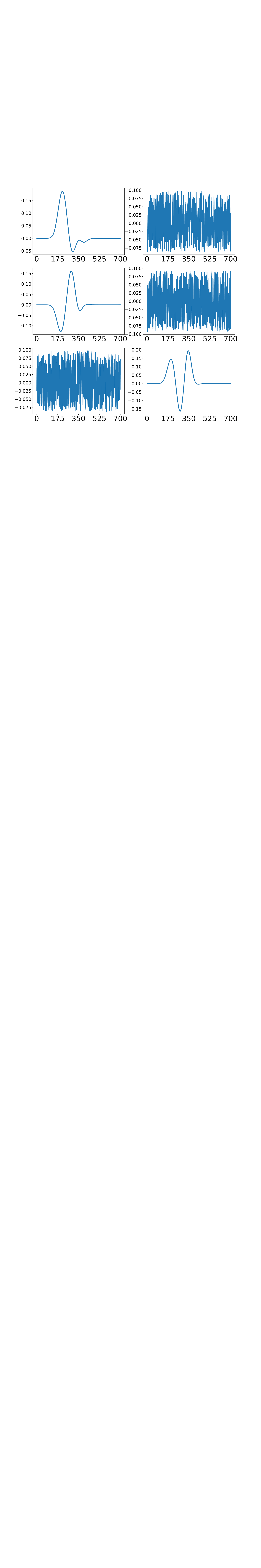}}
\subfigure[\ 8 neurons]{
\includegraphics[width=.45\columnwidth]{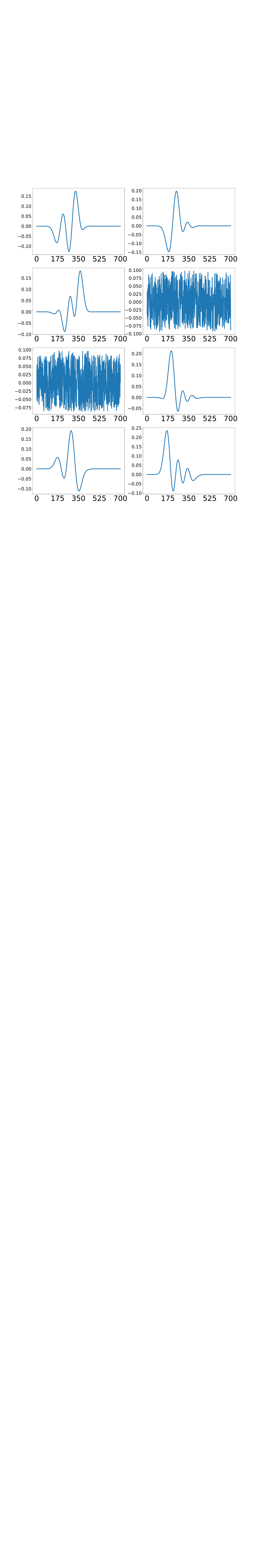}}
\caption{The extracted basis functions for machines with different neuron numbers in the NTO layer. Smooth curves are proper basis functions while noise-like curves indicate that the machine fails to find proper bases. The performance increases with the number of the proper bases so that the performance of the four-neuron case (b) is better than the six-neuron case (c).}
\label{basis}
\end{figure}

\begin{figure}[t]
\centering
\includegraphics[width=.9\columnwidth]{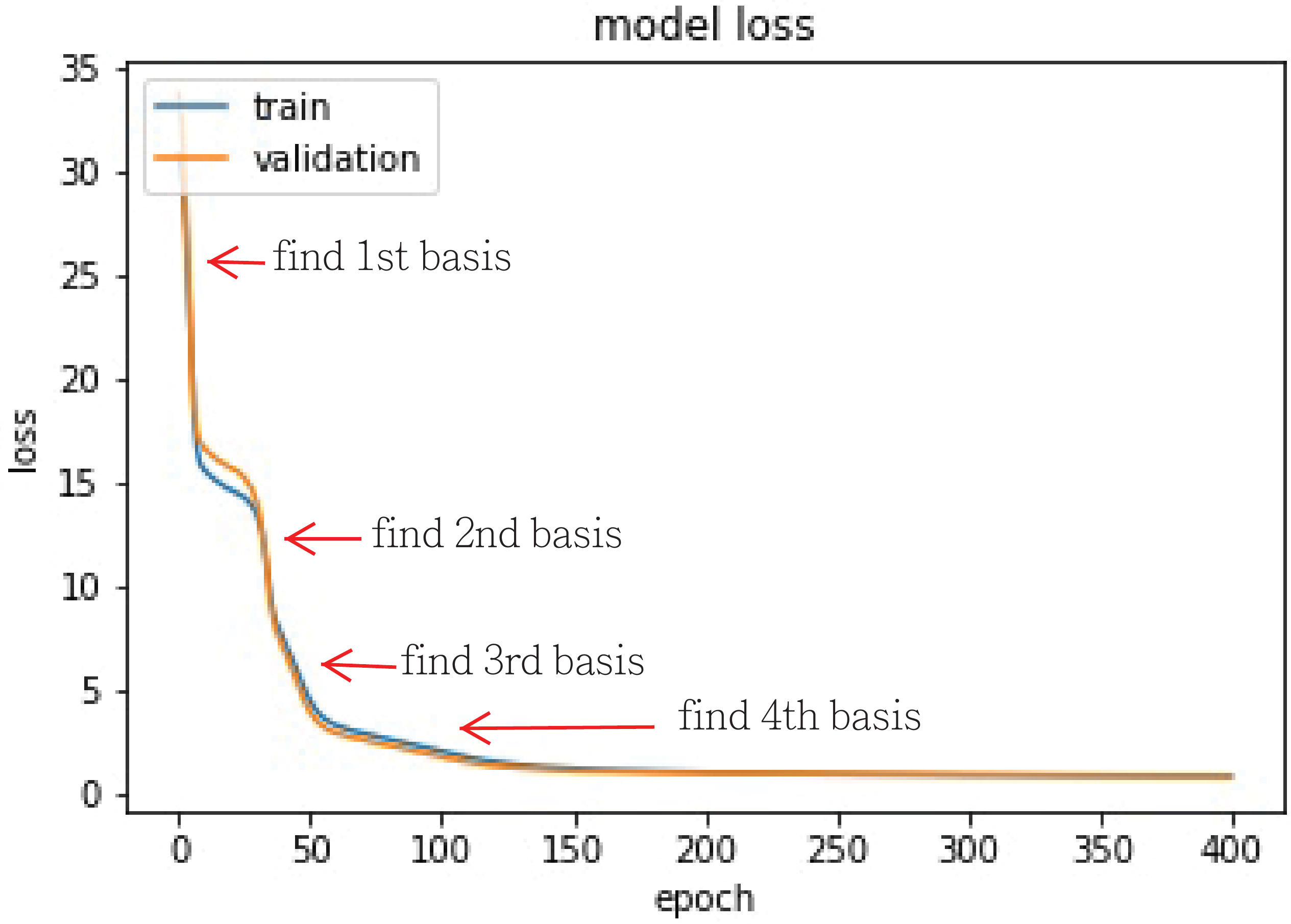}
\caption{The loss function for the four-neuron case during the training process. The loss function drops when the machine finds one basis function.}
\label{rmodel}
\end{figure}

First of all we analyze how the performance is influenced by the neuron number of the layer next to the output layer (NTO layer, see Fig. \ref{nn}). This neuron number turns out to be the number of basis functions whose combinations produce the output functions. We extract the weights between the NTO and output layers to find these bases. Suppose the output function $f(\omega_i)=\sum_j w_{ij}a_j$, where $a_j$ is the value of the $j$-th neuron of the NTO layer. The bias vector is ignored in this discussion as it does not influence the performance. Then $w_{ij}$ is the $i$-th element of the $j$-th basis. Note that the weights between the last two layers can be directly interpreted as bases because the activation function of the output layer is set to be linear, instead of nonlinear functions such as ReLU. We demonstrate cases when the NTO layer has different neuron numbers in Fig. \ref{basis} so that each curve corresponds to one basis function the machine has learned from data. Smooth curves are proper basis found by the machine while noise-like curves, such as the lower-left panel in Fig. \ref{basis}(c), indicate that the corresponding neuron is impotent. The more proper bases found by the machine, the better the performance. For instance, for the four-neuron case, the machine finds four proper bases so the performance is better than that in the six-neuron case where the machine only finds three proper bases, but worse than that in the eight-neuron case where the machine finds six proper bases. 
The connection between the performance and the number of proper basis can be further visualized from the training history. The four-neuron case is demonstrated in Fig. \ref{rmodel}. When a basis is found the loss function shows a drop and totally four proper bases (See Fig. \ref{basis}(b)) are found.

\section{Numerical Calculation of the Eliashberg equation}
\label{numerical}
The effects of the EPC are manifest in the mass renormalization factor $Z(\omega)$ and the gap function $\Delta(\omega)$, which can be determined theoretically by the material-dependent EBSF $\alpha^2 F(\omega)$. 
To have large enough datasets for the machine to learn, we generate various EBSFs randomly which suffice for our need of data. 
The ME equations (\ref{Eq:Ziw}) and (\ref{Eq:Zdelta}) are then solved self-consistently by inputting these generated EBSFs. 
Other quantities like $T$ and $\mu^*$ are input parameters to be specified to solve the equations.

The generated EBSF must be continuous and positive-definite. 
We first create random functions along the one-dimensional $\omega$ space by the geometric Brownian motion which is often used as the financial market price simulations \cite{GBM16, GBM14}.
The functions are then shifted to guarantee the positive-definite behaviour, and are further enforced to vanish at zero and large frequencies above the Debye cutoff.
The generated functions are then numerically smoothed to make themselves continuous. 
Finally, the functions can be mapped to the chosen low frequency region of interest.

In the whole work here, we map the generated EBSF to the frequency range of  $0 \sim 10$ meV, and set the temperature $T = 1.16 K$. The dimensionless empirical parameter $\mu^*$ is fixed to be 0.1 for simplicity since its value does not change the result much.
Note that the energy can be used as the common unit and other dimensional quantities are scaled correspondingly. For example, we can still get the same result of the gap function and renormalization if the frequency range is mapped to the range of $0 \sim 50$ meV, while T is chosen to be 5.8 $K$ simultaneously.
Some randomly-generated EBSFs will not produce superconductivity at the chosen temperature. We keep only the nonzero $\Delta$ data for training and testing.

In Eqs.\ (\ref{Eq:Ziw}) and (\ref{Eq:Zdelta}), the $Z$ and $\Delta$ are coupled all together with all Matsubara frequencies. 
In the numerical calculations of the coupled equations, the terms are truncated at high frequencies since $Z$ and $\Delta$ are decaying to fixed values at high Matsubara frequencies. The summation over different Matsubara frequencies $\omega_n$ is truncated to $n\in (-M-1, M)$ terms. The $M$ value determines the maximum Matsubara frequency included, which should be large enough to ensure convergent results, and especially high accuracies for the sake of analytical continuation to be performed later.

There always exists the solution $\Delta =0$ which corresponds to the normal state. In numerical calculation we should be careful of choosing the initial values and the convergent criteria to find the superconducting solutions ($\Delta \neq 0$) of interest.
To solve the coupled equations self-consistently, we solve them iteratively.
However, the simplest and straightforward iteration of a previous output as the next input usually converges poorly. 
Typically the trivial solution $\Delta=0$ might be obtained, especially when the non-trivial solution is small. 
To perform the numerical iteration properly, we adopt the Broyden's method, which is a quasi-Newton method to find roots of a set of nonlinear equations $\mathbf{f}(\mathbf{x})=0$ by updating the Jacobian matrix \cite{Broyden65}. Here $\mathbf{x} = \{ \Delta(i\omega_n), n = -(M+1), \ldots, M \}$ is the root solution to be found.
The Jacobian matrix $\mathbf{J}$ with $(i, j)$ entry $\mathbf{J}_{ij} = \frac{\partial f_i}{\partial x_j}$ specifies the first-order partial derivatives in each small displacement.
The $n-$th iteration is written in the form of 
\begin{equation}
\mathbf{J}_n \delta\mathbf{x}_n \simeq \delta \mathbf{f}_n,
\end{equation}
where $\mathbf{f}_n=\mathbf{f}(\mathbf{x}_n)$, $\delta\mathbf{f}_n = \mathbf{f}_n - \mathbf{f}_{n-1}$, and $\delta\mathbf{x}_n = \mathbf{x}_n - \mathbf{x}_{n-1}$. 
The inverse of the the Jacobian matrix is updated directly as
\begin{equation}
\mathbf{J}_n^{-1} = \mathbf{J}_{n-1}^{-1} + \frac{\delta\mathbf{x}_n - \mathbf{J}^{-1}_{n-1}\delta\mathbf{f}_n}
			{\delta\mathbf{x}_n^T\mathbf{J}^{-1}_{n-1}\delta\mathbf{f}_n} \delta \mathbf{x}_n^T \mathbf{J}^{-1}_{n-1}
\end{equation}
to minimize the Frobenius norm $|| \mathbf{J}_n - \mathbf{J}_{n-1}||_F$, as suggested by Broyden \cite{Broyden65}.
With $\mathbf{J}^{-1}$ determined, the roots of the equations are then improved by proceeding in the Newton direction:
\begin{equation}
\mathbf{x}_{n+1} = \mathbf{x}_n - \mathbf{J}^{-1}_n \mathbf{f}(\mathbf{x}_n)
\end{equation} 
until the solution converges.

Once the $Z$ and $\Delta$ are solved at the Matsubara frequencies, the analytical continuation of these functions to the real axis is then executed. We adopt the Pad{\'e} approximation, which requires sufficient and accurate N-points (here $2M+2$ points) data at the imaginary axis \cite{Pade}.
In practice, the truncated $M$ must be large enough to ensure the maximum Matsubara frequency several times of the real frequency range under discussion \cite{Carbotte90}.
According to our energy unit choice, our real frequency $\omega$ of interest is about $0 \sim 60$ meV, above which the $Z$ and $\Delta$ are usually saturated to 1 and 0, respectively. Only the low frequency parts below 35 meV of $Z$ and $\Delta$ are used as the input for the later machine learning. Our choice of $M = 350$ corresponds to a Matsubara frequency $\sim 220$ meV, about 22 times of the EBSF domain range and several times of the input frequency range, sufficient to resolve the low-frequency information.
If $Z(i\omega_n)$, $\Delta(i\omega_n)$, or $\phi(i\omega_n)$ do not converge for the chosen $M$ value, $M$ will be doubled until they converge. 
After these quantities are analytically continued to the real axis, we also compare $\Delta(\omega)$ and $\phi(\omega)/Z(\omega)$ 
for consistency check.
The two should be mathematically equal but may differ due to the adopted Pad{\'e} approximation or the numerical errors. We set the criteria to their difference and ratio to each other at each frequency point, and only those data having both criteria passed beyond $\omega= 35$ meV will be kept, since only the $Z(\omega)$ and $\Delta(\omega)$ below $35$ meV will be used for later machine learning to find reasonable relations among functions. 

\section{Determination of the appropriate latent dimension}
\label{properlatent}
For gap functions in the AE-generated $\{g_n\}$, the performance (of predicting the EBSF) can be solely determined by the loss function of the SL machine $S_n$, which is trained using $\{g_n\}$. How about gap functions other than the dataset $\{g_n\}$? We expect that for gap functions close enough to any member of $\{g_n\}$, the performance is also determined by the loss function of $S_n$ when $S_n$ is used to make predictions. Therefore, to determine the closeness of two data set, we can define a characteristic distance $l_n$ between two functions within $\{g_n\}$ and claim one gap function is close enough to $\{g_n\}$ when its closest distance to $\{g_n\}$ is less than $l_n$. For each member $g_{ni}$ in $\{g_n\}$, we can always find its closest partner with smallest distance $l_{ni}$ which is defined as the summation of the normalized absolute difference between two functions, $\sum_{\omega}|g_{ni}-g_{nj}|/\sum_{\omega}|g_{ni}|$, over all frequencies. Then $l_n$ is defined as the maximum of $\{l_{ni}\}$. We can then expect that one gap function not belonging to $\{g_n\}$ can share similar performance of $S_n$ if the distance between this gap function and its closest member in $\{g_n\}$, $l_{on}$, is less than $l_n$. We can then estimate the correct rate of the prediction by the portion of the original $\{g_o\}$ satisfying $l_{on}\leq l_n$ to further evaluate the performance of $\{g_o\}$ using $S_n$. In Table \ref{ratio}, we show this portion as well as the optimized value of the loss function of $S_n$ for different $d_L=n$.

\begin{table}[ht]
\centering
\caption{The portion of the original dataset included in the newly generated datasets with different latent dimensions ($d_L$'s). The optimized value of the loss function for each $d_L$ is also presented. For $d_L\geq 2$, the domains of newly generated datasets can cover almost all of the original dataset, which ensures the high correct rate of the predictions.}
\begin{tabular}{|c|c|c|c|c|c|c|}
\hline 
$d_L$ & 1 & 2 & 4 & 8 & 16 & 32 \\ 
\hline 
portion & $63.71\%$ & $99.39\%$ & $99.97\%$ & $99.97\%$ & $99.98\%$ & $99.99\%$ \\ 
\hline 
loss & $0.00001$ & $0.0001$ & $0.0038$ & $0.015$ & $0.025$ & $0.029$ \\ 
\hline
\end{tabular} 
\label{ratio}
\end{table}%

Please see Fig. \ref{aenewnew} to visualize the performance. The better the performance, the smaller the loss function. The correct rate is close to $100 \%$ for $d_L\geq 2$ and the performance decreases with increasing $d_L$. Therefore, the best choice will be $d_L=2$, consistent with na\"ive inspections of Fig. \ref{vaeall}.

\section{Relevant properties shared by original and newly generated datasets}
\label{relevantproperty}
\begin{figure}[t]
\centering
\includegraphics[width=.9\columnwidth]{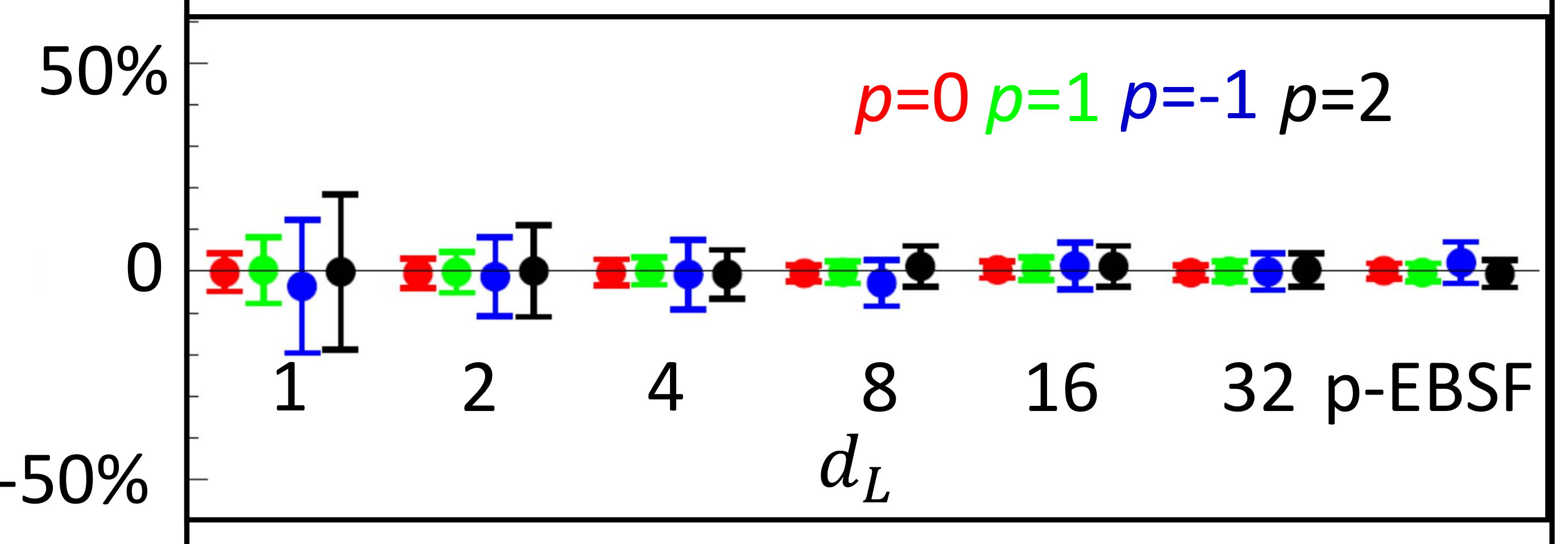}
\caption{The difference in different moments ($p$) between old (original) and newly SL-predicted EBSFs. We study the cases with $d_L=1,2,4,8,16,32$ as well as the p-EBSF. The error bars are standard deviations from the whole datasets. Statistically, differences in these moments between original and new EBSFs can be viewed as zeros, indicating that original and new EBSFs share key information about the gap functions.}
\label{moment}
\end{figure}

Since often superconducting properties, such as the critical temperature, can be estimated by some moments of the EBSFs without diving into the spectrum details \cite{allen1975,dynes1972,mcmillan1968}, we calculate the $p$-th moments $\mu_p = \int\omega^p f(\omega) d \omega$ carrying relevant information for comparison, where $f(\omega) = \alpha^2 F(\omega)$ and $p$ = 2, 1, 0, and -1 is calculated. Therefore, we can check if the original and SL-predicted EBSFs share the same properties. We calculate $\mu_{p,o}$ and $\mu_{p,n}$ which are $p$-th moment of old (original) and new (SL-predicted, $d_L=n$) EBSFs, respectively. These new EBSFs are obtained in three steps. At the first step, we obtain AE-transformed gap and EBSFs (See Appendix \ref{newdata}). At the second step, transformed gap functions and EBSFs are used to train a SL machine $S_n$. At the third step, we feed the old gap functions into $S_n$ to predict the new EBSFs and compare them to the old EBSFs. We then take the mean ($m$) and standard deviation ($std$) of the difference of $\frac{\mu_{p,o}-\mu_{p,n}}{\mu_{p,o}}$ over the whole dataset. The results, as shown in Fig. \ref{moment}, show that the old and new EBSFs do share relevant properties. We also compare the old EBSFs to the p-EBSF functions, which is consistent with Fig. \ref{pbsgtbs}.

\section{Generating new functions}
\label{newdata}
In this appendix, we provide detailed procedures of generating new EBSFs and gap functions for given $d_L=n$, which are used to train a SL machine. Consider one original gap function $g_o$ and EBSF $f_o$ as two vectors and we concatenate these two vectors to one combined vector $v_o$, the first part of which is the EBSF $f_o$ and the second part of which is the gap function $g_o$. We then train a AE with $d_L=n$, denoted as $A_n$, by using the dataset $\{v_o\}$ until the reconstruction error is optimized (or minimized). After the training is completed, we again feed $\{v_o\}$ to this trained AE and obtain the output $\{v_n\}$. Then the first part of each vector $v_n$ is the new EBSF $f_n$ and the second part of it is the new gap function $g_n$. Then we can use this dataset, $\{g_n\}$ and $\{f_n\}$, to perform the SL training mentioned in the main text.


\end{document}